\documentclass{npw}
\usepackage{graphicx}

\begin{document}

\title{An extended Agassi model: algebraic structure, phase diagram,
  and large size limit}
\author{
  JE Garc\'{\i}a-Ramos$^{1,2,3}$\email{enrique.ramos@dfaie.uhu.es}, J Dukelsky$^{4}$\email{dukelsky@iem.cfmac.csic.es},  P P\'erez-Fern\'andez$^{2,5}$\email{pedropf@us.es},  JM Arias$^{2,6}$\email{ariasc@us.es} \\
  $^1$\it\small Departamento de  Ciencias Integradas y 
  Centro de Estudios  Avanzados en F\'isica,\\
  \it\small Matem\'atica y Computaci\'on, Universidad de Huelva,
  21071 Huelva, Spain\\
  $^2$\it\small Instituto Carlos I de F\'{\i}sica Te\'orica y Computacional,\\
  \it\small Universidad de Granada, Fuentenueva s/n, 18071 Granada, Spain\\
  $^3$\it\small Unidad Asociada de la Universidad de Huelva al IEM (CSIC), Madrid,
  Spain\\
  $^4$\it\small Instituto de Estructura de la Materia, CSIC, Serrano 123, 28006
  Madrid, Spain\\
  $^5$\it\small Departamento de  F\'{\i}sica Aplicada III, Escuela T\'ecnica\\
  \it\small Superior de Ingenier\'{\i}a, Universidad de Sevilla, Sevilla,\\
  $^6$\it\small Departamento de F\'{\i}sica At\'omica, Molecular y Nuclear,\\
  \it\small Facultad de F\'{\i}sica, Universidad de Sevilla, Apartado~1065,
  41080 Sevilla, Spain}


\date{}

\maketitle

\begin{abstract}
  The Agassi model \cite{Agas68} is a schematic two-level model that involves pairing and monopole-monopole interactions. It is, therefore, an extension of the well known Lipkin-Meshkov-Glick (LMG) model \cite{Lipk65}. In this paper we review the algebraic formulation of an extension of the Agassi model as well as its bosonic realization through the Schwinger representation. Moreover, a mean-field approximation for the model is presented and its phase diagram discussed. Finally, a $1/j$ analysis, with $j$ proportional to the degeneracy of each level, is worked out to obtain the thermodynamic limit of the ground state energy and some order parameters from the exact Hamiltonian diagonalization for finite$-j$.
\end{abstract}
  

\section{Introduction}
\label{sec-intro}

Symmetry plays a major role in physics to handle problems that because of its complexity or size cannot be treated by \textit{brute force} diagonalizations or with standard many-body approximations. On the other hand, algebraic boson or fermion models with few-degrees of freedom are easily solved and have provided for years with benchmarks to test many-body approximations. These models are characterized by Lie algebras \cite{Iach06} and for particular situations (called dynamical symmetries) can be solved analytically. In addition, in more general situations they can be solved numerically even for very large system sizes. In different areas of physics these families of models can be found. For instance, in quantum optics popular models for radiation-matter interaction are the Jaynes-Cummings \cite{Jayn63} or the Dicke \cite{Dick54} models. In nuclear physics, the interacting boson model (IBM) \cite{Iach87} is based on the su(6) and has been used extensively in the last years for the study of medium and heavy nuclei. Also in nuclear physics the Elliot SU(3) model \cite{Elli58} was proposed to study rotational structures in nuclei. Other simple two-level models as the LMG model \cite{Lipk65} and the two-level pairing model \cite{Hoga61} were originally proposed in the context of nuclear physics but then their simplicity and success allow them to be used in different Physics branches: condense matter, molecular physics or cold atom physics. 

The Agassi model was proposed in the late 1960's by D.\ Agassi as a model that could mimic the much more complex pairing-plus-quadrupole model of nuclear physics \cite{Agas68}. To this end, he tailored a model through the interplay of a two-level pairing plus a monopole term. In this sense, the Agassi model is an extension of the LMG model which, in addition, contains a pairing interaction and, therefore, is a much more rich model. The model was proposed originally to test the goodness of two popular many-body approximations, namely, the Bardeen-Cooper-Schrieffer (BCS) and the Hartree-Fock-Bogoliubov (HFB) approximations. The Agassi model allows its exact diagonalization, that can be carried out easily owing the algebraic structure of the model \cite{Agas68}. These exact results can be confronted to those obtained with approximation techniques to validate under which conditions these approximations are sufficiently precise. In spite of its unquestionable attractive, the Agassi model has been seldom used for years regardless of its great flexibility and its simplicity to  be solved for large systems. There have been few works published in which the random phase approximation (RPA), HFB \cite{Agas68,Schu69, Davi86} or perturbation theory \cite{Gert83} were applied to the Agassi model. Moreover, modern many-body approximations that have been used intensively in nuclear physics did not take advantage of the model to check their applicability, and to establish their limitations and accuracy. Only very recently, an application to explore the coupled cluster theory in relation to symmetry breaking and restoration \cite{Herm17} has been published. This work opens the possibility of future applications to quantum chemistry and nuclear physics. 

In addition, the Agassi model can play a relevant role in the field of
quantum phase transitions (QPTs). The study of QPTs  and critical  points in algebraic models have received much attention in the last two decades. Many publications on the subject have been done (see, e.g.,
\cite{Cejn09} and  \cite{Cejn10}). In particular, QPTs have been
studied within the IBM \cite{Gino80,Feng81}, the LMG \cite{Vida06},
the vibron model \cite{Iach08}, and the Jaynes-Cummings \cite{Jayn63}
and Dicke \cite{Dick54} models. Therefore, it seems quite natural to
use the Agassi model, which encompasses the LMG and the two-level
paring models, with an underlying so(5) algebra, to study the QPTs
that arise in it. In Ref. ~\cite{Davi86} the Agassi model phase diagram was established. This diagram presents, apart from the symmetric phase, two broken symmetry phases: one connected to parity breaking related to the monopole interaction and another, called
superconducting phase, linked to the pairing interaction. Recently, an
extended Agassi model, in which a more general monopole interaction is
included, was proposed in Ref. \cite{Garc18}. The corresponding richer
phase diagram was reported and it includes several QPTs of different
character. The model was studied using the mean-field HFB theory and the results checked against exact diagonalizations.  

In this paper, the extended Agassi model is revised and its boson
image using the Schwinger representation is worked out. Then, the phase diagram of the model is discussed and a technique that allows to
obtain the mean field results as the large size limit of exact finite number of particle calculations is introduced. The paper is organized as follows: in Section \ref{sec-model}, the algebraic structure of the
proposed extended Agassi model is discussed. Also in Section \ref{sec-model}, the Schwinger representation of the model is worked out. In Section \ref{sec-mf}, a mean-field approach based on a condensate of bosons plus the HFB approach is applied to obtain the corresponding mean-field energy surfaces and to analyze their stability. In this section, the phase diagram structure and the nature of the QPTs are presented. In Section \ref{sec-large}, a $1/j$ analysis of several quantities is proposed to obtain their thermodynamic limit. Finally, Section \ref{sec-conclusions} is devoted to the summary and conclusions of this work. 

\section{Algebraic structure of Agassi model} 
\label{sec-model}

The Agassi model is a two-level system, each of the levels has $\Omega$ (even) degeneracy. The labeling for the single particle states is the following: $\sigma=1$ for the upper level and $\sigma=-1$ for the lower one, plus an additional quantum number $m=\pm 1, \pm 2, ..., \pm \Omega/2$, which labels the states within a given level. Consequently, one can use a spin image taking $j$ as an integer number and $\Omega=2j$. Moreover, $\sigma$ is used as the level parity, $\sigma=+1$ and $\sigma=-1$ imply positive and negative parity, respectively. 
The extended Agassi model Hamiltonian is,
\begin{eqnarray}
H=\varepsilon J^{0} &-& g\sum_{\sigma,\sigma^{\prime}=-1,1}A_{\sigma}^{\dagger}
     A_{\sigma^{\prime}} \nonumber \\
  &-&\frac{V}{2}\left[  \left(  J^{+}\right)  ^{2}+\left(
J^{-}\right)  ^{2}\right]  -2hA_{0}^{\dagger}A_{0}.
\label{eq_h_agassi}
\end{eqnarray}
All operators in the Hamiltonian (\ref{eq_h_agassi}) are defined in terms of the fermion creation and annihilation operator as,
\begin{eqnarray}
          \label{J}
J^{+}&=&\sum_{m=-j}^{j}c_{1,m}^{\dagger}c_{-1,m}=\left(  J^{-}\right)  ^{\dagger},\\
  J^{0}&=&\frac{1}{2}\sum_{m=-j}^{j}\left(c_{1,m}^{\dagger}c_{1,m}%
           -c_{-1,m}^{\dagger}c_{-1,m}\right) ,
\end{eqnarray}
%
\begin{eqnarray}
  A_{1}^{\dagger}&=&\sum_{m=1}^{j}c_{1,m}^{\dagger}c_{1,-m}^{\dagger},\\
  A_{-1}^{\dagger}&=&\sum_{m=1}^{j}c_{-1,m}^{\dagger}c_{-1,-m}^{\dagger},\\
  A_{0}^{\dagger}&=&\sum_{m=1}^{j}\left(  c_{-1,m}^{\dagger}c_{1,-m}^{\dagger
                       }-c_{-1,-m}^{\dagger}c_{1,m}^{\dagger}\right) ,
\label{eq-As}
\end{eqnarray}
\begin{eqnarray}
  A_{1}&=&\sum_{m=1}^{j}c_{1,-m}c_{1,m},\\
  A_{-1}&=&\sum_{m=1}^{j}c_{-1,-m}c_{-1, m},\\
  A_{0}&=&\sum_{m=1}^{j}\left(c_{1,-m}c_{-1, m}-c_{1,m}c_{-1,-m}\right),
\label{eq-As-herm}
\end{eqnarray}
\begin{equation}
N_{\sigma}=\sum_{m=-j}^{j}c_{\sigma, m}^{\dagger}c_{\sigma, m},\qquad
N=N_{1}+N_{-1},%
\label{N}
\end{equation}
where $c^\dagger_{\sigma, m}$, $c_{\sigma, m}$ are single fermion creation and annihilation operators in the state ${|\sigma, m\rangle}$, respectively. 
The last term in (\ref{eq_h_agassi}), $-2hA_{0}^{\dagger}A_{0}$, has been added to the original Agassi model. This term does not correspond to the standard pairing since it involves correlated pairs with opposite angular momentum projection, but sitting in different levels, while the standard one treats with correlated pairs sitting in the same level, either the lower or the upper one. The addition of this extra term gives rise to new relevant effects not present in the original model formulation.
It is convenient to redefine the Hamiltonian parameters to have them rescaled  with the shell size. Thus, new parameters $\chi$, $\Sigma$, and $\Lambda$ (see \cite{Davi86}) are introduced as,
\begin{equation}
V=\frac{\varepsilon\chi}{2j-1},\quad g=\frac{\varepsilon\Sigma}{2j-1},\quad
h=\frac{\varepsilon\Lambda}{2j-1}.%
\label{eq_par}\end{equation}
These three model parameters have to be positive in order to do not produce unphysical situations. With this parametrization, the extended Agassi Hamiltonian is,
\begin{eqnarray}
  \nonumber
  H&=&\varepsilon \left[ J^{0}- \frac{\Sigma}{2j-1}~\sum_{\sigma\sigma^{\prime}}A_{\sigma}^{\dagger}
       A_{\sigma^{\prime}}\right.\\
  \nonumber
  &-&\left.\frac{\chi}{2 (2j-1)}~\left[  \left(  J^{+}\right)  ^{2}+\left(
      J^{-}\right)  ^{2}\right]\right.  \\
  &-&\left. 2
      \frac{\Lambda}{2j-1}~A_{0}^{\dagger}A_{0}\right]. 
\label{eq_h_agassi-si}
\end{eqnarray}
\begin{figure}[hbt]
  \centering
\includegraphics[width=1\linewidth]{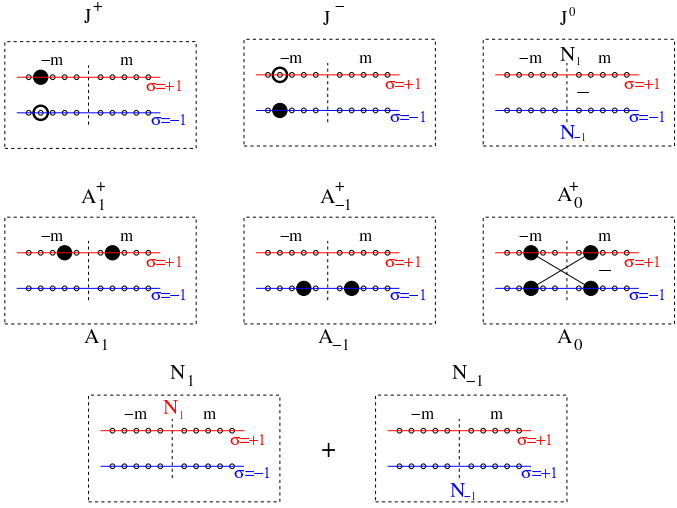}
\caption{Schematic representation of the O(5) generators that underlay the extended Agassi model. For the $J$ operators the full dot represents the creation of a fermion while the open dot represents the annihilation. For the $A^\dagger$ operator, the full dot represents the creation of a fermion while for the $A$ operators represents the annihilation. $N_1$ ($N_{-1}$) counts the number of fermions in the $\sigma=+1$ ($\sigma=-1$) level.}
\label{fig-generators}
\end{figure}
The model algebraic structure is o(5), with $10$ independent generators: six $A$'s, plus three $J$'s, plus the particle number, $N$. It is worth noting that $N_{1}$ and $N_{-1}$ are linear combinations of $J^{0}$ and $N$. In a pictorial way the operators of the model are represented in Fig.~\ref{fig-generators}. It is of interest to write down the commutation relations of the operators (\ref{J})-(\ref{N}) that underlie an o(5) algebra,
\begin{equation}
\left[  J^{+},J^{-}\right]  =2J^{0},\quad\left[  J^{0},J^{\pm}\right]  =\pm
J^{\pm},
\end{equation}
\begin{eqnarray}
  \left[  A_{1},A_{1}^{+}\right]  &=&  j-N_{1},\\
  \left[  A_{-1},A_{-1}^{+}\right]  &=&  j-N_{-1}, \\
  \left[  A_{0},A_{0}^{+}\right]  &=&  2j-N,
\end{eqnarray}
\begin{equation}
\left[  A_{0},A_{1}^{+}\right]  =-J^{+},\qquad \left[  A_{0},A_{-1}^{+}\right]
=-J^{-},
\end{equation}
\begin{equation}
\left[  A_{1},A_{0}^{+}\right]  =-J^{-},\qquad \left[  A_{-1},A_{0}^{+}\right]
=-J^{+},
\end{equation}
\begin{equation}
\left[  J^{+},A_{1}^{+}\right]  =0,~ \left[  J^{+},A_{-1}^{+}\right]
=A_{0}^{+},~\left[  J^{+},A_{0}^{+}\right]  =2A_{1}^{+},
\end{equation}
\begin{equation}
\left[  J^{-},A_{1}\right]  =0,~\left[  J^{-},A_{-1}\right]  =-A_{0},~\left[
J^{-},A_{0}\right]  =-2A_{1},
\end{equation}
\begin{equation}
\left[  J^{+},A_{1}\right]  =-A_{0},~\left[  J^{+},A_{0}\right]
=-2A_{-1},~\left[  J^{+},A_{-1}\right]  =0 ,
\end{equation}
\begin{equation}
\left[  J^{-},A_{1}^{+}\right]  =A_{0}^{+},~\left[  J^{-},A_{0}^{+}\right]
=2A_{-1}^{+},~\left[  J^{-},A_{-1}^{+}\right]  =0,
\end{equation}
\begin{equation}
\left[  J^{0},A_{1}^{+}\right]  =A_{1}^{+},~\left[  J^{0},A_{-1}^{+}\right]
=-A_{-1}^{+},~\left[  J^{0},A_{0}^{+}\right]  =0,
\end{equation}
\begin{equation}
\left[  J^{0},A_{1}\right]  =-A_{1},~\left[  J^{0},A_{-1}\right]
=A_{-1},~\left[  J^{0},A_{0}\right]  =0.
\end{equation}
Therefore, the Hamiltonian (\ref{eq_h_agassi-si}) can be diagonalized within a o(5) basis \cite{Agas68,Hirs97}. On the other hand, the system eigenstates have a well defined parity since the parity operator $e^{-\imath \pi J^{0}}$ commutes with the Hamiltonian (\ref{eq_h_agassi-si}).
The complete, but non-orthogonal, basis of the model is defined in terms of three quantum numbers: $n_+$ for the number of pairs in the upper orbit ($\sigma=1$), $n_-$ for the number of pairs in the lower orbit ($\sigma=-1$), and $n_0$ for the number of  $A_0$ pairs
\begin{equation}
\left\vert n_+,n_-,n_0\right\rangle =  \left(A_{1}^{+}\right)^{n_+} \left(A_{-1}^{+}\right)^{n_-} \left(A_{0}^{+}\right)^{n_0}\left\vert
0\right\rangle.
\end{equation}
In order to perform calculations within the Agassi model, one has to select the $j$ quantum number, that fixes the system size, and the number of fermions, $N$ in the space. Here, the ratio $N/4j$ is fixed to the value $1/2$ ($N=2j$) which means half-filling or, equivalently, $j$ fermion pairs are considered. In this  half-filling situation, the quantum numbers fulfill $n_-+n_++n_0=j$.

\subsection{Schwinger representation}%
\label{sec-schwinger}

It is of interest to build the boson image of the extended Agassi model through the Schwinger representation. To this end one introduces five bosons: $p^\dagger_{-1}$ which creates a correlated fermion pair in the lower level ($\sigma=-1$),  $p^\dagger_{1}$ which creates a correlated fermion pair in the upper level ($\sigma=1$), $p^\dagger_{0}$ which creates a correlated fermion pair with a given $m$ (one fermion in the upper level and the other in the lower one), $v^+$ which creates a quartet of fermions (one pair $(m,-m)$ above and other pair $(m,-m)$ below), and $u^+$ that corresponds to the identity (vacuum). These five operators ($b^\dagger_i\equiv p^\dagger_{-1}, p^\dagger_{0}, p^\dagger_{1}, u^\dagger, v^\dagger$)  fulfill the
usual boson commutation relations,
\begin{equation}
  [b_i,b^\dagger_j]=\delta_{i j},~ [b_i^\dagger,b^\dagger_j]=0,~[b_i,b_j]= 0.
\end{equation} 

The O(5) generators in terms of Schwinger representation are written as,
\begin{eqnarray}
  J^{0}&=&p_{1}^{\dagger}p_{1}-p_{-1}^{\dagger}p_{-1},\\
  J^{+}&=&\sqrt{2}\left(p_{1}^{\dagger}p_{0}+p_{0}^{\dagger}p_{-1}\right),\\
  J^{-}&=&\sqrt{2}\left(p_{-1}^{\dagger}p_{0}+p_{0}^{\dagger}p_{1}\right),
\end{eqnarray}

\begin{eqnarray}
  A_{1}^{\dagger}&=&p_{1}^{\dagger}u+v^{\dagger}p_{-1},\\
  A_{0}^{\dagger}&=&\sqrt{2}\left(
                     p_{0}^{\dagger}u-v^{\dagger}p_{0}\right),\\
  A_{-1}^{\dagger}&=&p_{-1}^{\dagger}u+v^{\dagger}p_{1}
\end{eqnarray}

\begin{eqnarray}
  A_{1}&=&p_{-1}^{\dagger}v+u^{\dagger}p_{1},\\
  A_{0}&=&\sqrt{2}\left(u^{\dagger}p_{0}-p_{0}^{\dagger}v\right),\\
  A_{-1}&=&p_{1}^{\dagger}v+u^{\dagger}p_{-1},
\end{eqnarray}

\begin{eqnarray}
  N_{1}&=&p_{0}^{\dagger}p_{0}+2p_{1}^{\dagger}p_{1}+2v^{\dagger}v,\\
  N_{-1}&=&p_{0}^{\dagger}p_{0}+2p_{-1}^{\dagger}p_{-1}+2v^{\dagger}v,\\
  N&=& N_{-1}+ N_{1}.
\end{eqnarray}

The Hamiltonian (\ref{eq_h_agassi}) in terms of Schwinger bosons is written as
\begin{eqnarray}
  \nonumber
 &&H_{B}    =\left(  \varepsilon-g\right)  p_{1}^{\dagger}p_{1}-\left(
               \varepsilon+g\right)  p_{-1}^{\dagger}p_{-1}\\
  \nonumber
 &-&\left(  g+V\right)  \left(
             p_{1}^{\dagger}p_{-1}+p_{-1}^{\dagger}p_{1}\right)
             -2gv^{\dagger}v -2h \left(p_0^\dagger p_0+v^\dagger v \right)\\
 &-&V\left(  p_{1}^{\dagger}p_{1}^{\dagger}p_{0}p_{0}+p_{-1}^{\dagger}%
               p_{-1}^{\dagger}p_{0}p_{0}\right.\nonumber \\
 &+& \left.p_{0}^{\dagger}p_{0}^{\dagger}p_{1}p_{1}%
             +p_{0}^{\dagger}p_{0}^{\dagger}p_{-1}p_{-1}+2p_{1}^{\dagger}p_{0}^{\dagger
             }p_{0}p_{-1}+2p_{-1}^{\dagger}p_{0}^{\dagger}p_{0}p_{1}\right)
             \nonumber \\
&-& g\left[  2p_{1}^{\dagger}p_{-1}^{\dagger}uv+2u^{\dagger}v^{\dagger}%
               p_{-1}p_{1}\right. \nonumber \\
&+& \left.p_{1}^{\dagger}u^{\dagger}up_{1}+p_{1}^{\dagger}v^{\dagger}%
             vp_{1}+p_{-1}^{\dagger}u^{\dagger}up_{-1}+p_{-1}^{\dagger}v^{\dagger}%
             vp_{-1}\right. \nonumber \\
&+& \left.  p_{1}^{\dagger}p_{1}^{\dagger}uv+u^{\dagger}v^{\dagger}p_{1}%
             p_{1}+p_{-1}^{\dagger}p_{-1}^{\dagger}uv+u^{\dagger}v^{\dagger}p_{-1}p_{-1}\right.\nonumber \\
&+& \left.p_{1}^{\dagger}u^{\dagger}up_{-1}+p_{1}^{\dagger}v^{\dagger}
             vp_{-1}
             + p_{-1}^{\dagger}u^{\dagger}up_{1}+p_{-1}^{\dagger}v^{\dagger}vp_{1}\right] \nonumber\\
&-& 2h \left(p_0^\dagger u^\dagger u p_0- p_0^\dagger
             p_0^\dagger uv- u^\dagger v^\dagger p_0 p_0 +v ^\dagger
             p_0^\dagger p_0 v \right).  
             \label{H_schwinger}
\end{eqnarray}

In order to diagonalize the above Hamiltonian the following basis can be used,
\begin{equation}
\left|n_{-1},n_0,n_1,n\right\rangle=
\frac{(p_{-1}^\dag)^{n_{-1}}(p_0^\dag)^{n_0}(p_1^\dag)^{n_1}(u^\dag)^n (v^\dag)^n}{\sqrt{ n_{-1}! n_0! n_1! (n!)^2}}|0\rangle.
\label{basis-schwinger}
\end{equation}
The states should fulfill the number constraint, which for the half-filled case implies
\begin{equation}
n_{-1}+n_0+n_{1}+2~n=j.
\end{equation}
Because $n$ is an integer, $j-n_0-n_1-n_{-1}$ should be an even number. In addition, the states own a given parity, positive for even values of $n_0$ and negative for odd values.

The space spanned by the basis states (\ref{basis-schwinger}) corresponds to a representation of the su(5) algebra. However, we are interested in the so(5) subspace that has a one to one correspondence with the physical states of the fermionic Agassi model. This physical subspace is characterized by a null value of the quadratic Casimir operator of so(5) that can be written as a boson pairing interaction,
\begin{equation}
P^{\dagger}P=\left(  2p_{1}^{\dagger}p_{-1}^{\dagger}-p_{0}^{\dagger}%
p_{0}^{\dagger}-2u^{\dagger}v^{\dagger}\right)  \left(  2p_{1}p_{-1}%
-p_{0}p_{0}-2uv\right),
\end{equation}
meaning that these states have a generalized seniority zero.  Alternatively, this condition can be used to select the physical states after a diagonalization in the enlarged basis (\ref{basis-schwinger}) as those eigenstates that fulfill,
\begin{equation}
P\left\vert \Phi\right\rangle =\left(  2p_{1}p_{-1}-p_{0}p_{0}-2uv\right)
\left\vert \Phi\right\rangle =0 .
\label{constraint1}
\end{equation}


\section{Mean-field approach for the extended Agassi model}
\label{sec-mf}
In this section, the extended Agassi model is studied by mean-field techniques. Two different approaches that lead to the same results are elaborated. First, a boson condensate depending on some variational parameters is used to  obtain the energy surface associated to the Hamiltonian within its Schwinger image and, second, the HFB approach with the fermionic image of the Hamiltonian is worked out. 

\subsection{Mean field with a boson condensate in the Schwinger representation}
\label{sec-mf-sch}
The initial point of this variational method is a boson condensate, $\Gamma^\dag$, that is proposed to be a linear combination of the five bosons appearing in the Schwinger image and five variational parameters, namely, $\eta_{-1}$, $\eta_0$, $\eta_1$, $\delta$, and $\gamma$,
\begin{equation} 
  \Gamma^\dagger=\frac{\eta_{-1} p_{-1}^\dagger+ \eta_{0}p_{0}^\dagger
    +\eta_{1}p_{1}^\dagger+ \delta u^\dagger+\gamma
    v^\dagger}{\sqrt{\eta_{-1}^2+ \eta_{0}^2+ \eta_{1}^2+
        \delta^2+\gamma^2}}.
\end{equation} 
Then, the trial wave function of the system ground state is proposed to be a condensate of j boson pairs as,
\begin{equation} 
  | j, \eta_{-1}, \eta_{0}, \eta_{1}, \delta,\gamma  \rangle =
  \frac{\left(\Gamma^\dagger\right)^j}{\sqrt{j!}}|0\rangle .
  \label{condensate}
\end{equation} 
Note that it is assumed that the number of fermion pairs is $j$ and, therefore, the system is half filled.

Next step is to calculate the expectation value of the Hamiltonian (\ref{H_schwinger}) with the trial wave function (\ref{condensate})
\begin{eqnarray}
  \lefteqn{E(j,\eta_{-1}, \eta_{0}, \eta_{1}, \delta,\gamma )}\nonumber\\
  & & = \langle j,
  \eta_{-1}, \eta_{0}, \eta_{1}, \delta,\gamma | H  | j, \eta_{-1},
  \eta_{0}, \eta_{1}, \delta,\gamma  \rangle ,
\end{eqnarray}
to obtain the corresponding energy surface. In the $j\rightarrow\infty$ limit this energy surface reads as,
\begin{eqnarray}
\nonumber E &=& \varepsilon j\frac{\eta_{-1}^2+\eta_{1}^2}{\eta_{-1}^2+ \eta_{0}^2+ \eta_{1}^2+
                                     \delta^2+\gamma^2} \\
  \nonumber &-& g j^2 \frac{(\eta_{-1}^2+\eta_{1}^2)(\delta+\gamma)^2}{\left(\eta_{-1}^2+ \eta_{0}^2+ \eta_{1}^2+
                \delta^2+\gamma^2\right)^2}\\
  \nonumber &-& V j^2\frac{2(\eta_{-1}+\eta_{1})^2 \eta_0^2} {\left(\eta_{-1}^2+ \eta_{0}^2+ \eta_{1}^2+
                \delta^2+\gamma^2\right)^2}\\
            &-& 2 h j^2 \frac{\eta_0^2(\delta-\gamma)^2 } {\left(\eta_{-1}^2+ \eta_{0}^2+ \eta_{1}^2+
                \delta^2+\gamma^2\right)^2},
\end{eqnarray}
where the contributions from the two-body terms to the one-body terms have been neglected, since they vanish assuming that the two-body coefficients are scaled with an extra $1/j$ factor. As already commented before, the use the parameters (\ref{eq_par}) that take into account the sistem size $j$ is convenient. Using them, the energy functional is, 
\begin{eqnarray}
  \nonumber \frac{E}{\varepsilon j} &=& \frac{\eta_{-1}^2+\eta_{1}^2}{\eta_{-1}^2+ \eta_{0}^2+ \eta_{1}^2+
                                        \delta^2+\gamma^2} \\
  \nonumber &-& \frac{\Sigma}{2} \frac{(\eta_{-1}^2+\eta_{1}^2)(\delta+\gamma)^2}{\left(\eta_{-1}^2+ \eta_{0}^2+ \eta_{1}^2+
                \delta^2+\gamma^2\right)^2}\\
  \nonumber &-& \chi\frac{(\eta_{-1}+\eta_{1})^2 \eta_0^2} {\left(\eta_{-1}^2+ \eta_{0}^2+ \eta_{1}^2+
                \delta^2+\gamma^2\right)^2}\\
                                    &-& \Lambda \frac{\eta_0^2(\delta-\gamma)^2 } {\left(\eta_{-1}^2+\eta_{0}^2+ \eta_{1}^2+
                                        \delta^2+\gamma^2\right)^2}.
                \label{E_sch}
\end{eqnarray}
Note that $\varepsilon$ is an overall energy constant, that in practice can be considered as $\varepsilon=1$. Please note, that in mean-field a large sistem size $j$ is assumed, consequently, the term $-1$ in the denominator $(2j-1)$ of Eqs. (\ref{eq_par}) is negligible.

In principle, the energy surface depends on five variational parameters, but in practice, this number is smaller. First, for positive values of $\varepsilon$, which corresponds to the physical cases, $\eta_{-1}$ can be fixed to  $\eta_{-1}=1$. Moreover, because of the shape of the function (\ref{E_sch}), either $\delta=\gamma$ or $\delta=-\gamma$ for all possible minima. This fact has very important consequences on the way the energy function behaves with respect to the control parameters. As a matter of fact, parameters $\Sigma$ and $\Lambda$ never affect the value of the energy in its minimum at the same time. In general, for small enough values of $\Lambda$ the energy only depends on $\chi$ and $\Sigma$, while for large enough values of $\Lambda$ the energy only depends on $\chi$ and $\Lambda$. Therefore, the system behaves as if two separated energy surfaces were competing. One has to look for the minima of the two competing surfaces by minimizing both energy surfaces with respect to the variational parameters and look for the global minimum. The detailed study of the energy surfaces in the Schwinger representation can be carried out both numerical and analytically, however, the results are fully equivalent to those obtained with the HFB approach. Since the HFB derivation is more straightforward, only this approach is worked out in detail in the next subsection. 

\subsection{The Hartree-Fock-Bogoliubov approach}
\label{sec-hfb}

\begin{figure}[hbt]
  \centering
\includegraphics[width=.9\linewidth]{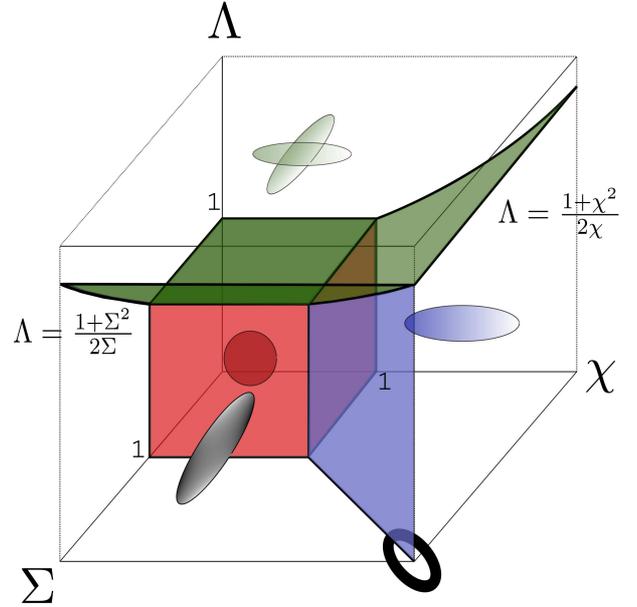}
\caption{Graphical representation of the phase diagram of the extended Agassi Hamiltonian
  (\ref{eq_h_agassi-si}). Red vertical planes represent second
  order QPT surfaces. The green surface ($\Lambda=1$ for $\chi<1$ and
  $\Sigma<1$, $\Lambda=\frac{1+\chi^2}{2\chi}$ for $\chi>\Sigma$ and
  $\Lambda=\frac{1+\Sigma^2}{2\Sigma}$ for $\chi<\Sigma$) and the blue
  vertical one ($\chi=\Sigma$ and $\Lambda <
  \frac{1+\Sigma^2}{2\Sigma}$) correspond to first order critical
  surfaces. Red sphere, blue oval, black oval, black thick oval, and
  crossed green ovals correspond to the symmetric solution, the HF
  deformed solution, the BCS deformed solution, the closed valley solution, and HF-BCS deformed solution, respectively. Figure adapted from Ref.~\cite{Garc18}.} 
\label{fig-phase-dia}
\end{figure}

Our mean-field  treatment for the extended Agassi model follows
closely the approach presented in \cite{Davi86,Garc18}. First, a Hatree-Fock transformation is applied, and then it is complemented with a Bogoliubov one.
Different energy surfaces, called A and B, are obtained with two alternative phase selections for the Bogoliubov transformation (see \cite{Garc18} for details).
The variational mean-field state is defined via the Hatree-Fock-Bogoliubov (HFB) formalism in terms of the parameters $\varphi$ and $\beta$. Then, the expectation value of the Hamiltonian (\ref{eq_h_agassi-si}) produces an energy surface, 
\begin{equation}
  E(\varphi, \beta) = \frac{\langle HFB(\varphi, \beta)| H  | HFB(\varphi, \beta) \rangle}{\langle HFB(\varphi, \beta)|HFB(\varphi, \beta) \rangle}.
\end{equation}
The minimization with respect to $\varphi$ and $\beta$ of the energy functional $E(\varphi, \beta)$ allows one to find the extrema. Their characterization (maximum, minimum or saddle point) is done by constructing the Hessian matrix. The complete analysis for the two alternative phase selections is done in Ref.~\cite{Garc18}, a short summary is presented here for completeness. Depending on the phase selection two energy surfaces, A and B, are obtained with scaled energies,
\begin{eqnarray}
\frac{E_{A}}{j\varepsilon}&=&-\cos\varphi\cos\beta-\frac{\Sigma}{2}\sin^{2}%
\beta-\frac{\chi}{2}\sin^{2}\varphi\cos^{2}\beta.
\label{eq-energ1b} \\
\frac{E_{B}}{j\varepsilon}&=&-\cos\varphi\cos\beta-\Lambda\sin^{2}\beta\sin
                              ^{2}\varphi-\frac{\chi}{2}\sin^{2}\varphi\cos^{2}\beta.\nonumber\\
 ~&~  
\label{eq_energ2b}
\end{eqnarray}
In order to determine the ground state of the system one has to look for the absolute minima of the two surfaces. Because of the existence of two surfaces there are regions in which different phases coexist. The detailed discussion on these points was presented in Ref. \cite{Garc18}. Five different phases are established in the phase diagram (Fig.~\ref{fig-phase-dia})

\begin{itemize}
\item Symmetric or spherical solution: it is the region with $\chi <1$, $\Sigma<1$ and $\Lambda<1$.

\item HF deformed solution: it is the region $\chi>1$, $\chi>\Sigma$ and $\Lambda< \frac{1+\chi^2}{2 \chi}$. 

\item BCS deformed solutions: it is the region $\Sigma>1$, $\Sigma>\chi$ and  $\Lambda< \frac{1+\Sigma^2}{2 \Sigma}$. 

\item Combined HF-BCS deformed solution: it is the region with $\Lambda>1$, $\Lambda> \frac{1+\Sigma^2}{2 \Sigma}$ and  $\Lambda>\frac{1+\chi^2}{2 \chi}$.  

\item Closed valley solution: it is the plane $\chi=\Sigma$ with $\Lambda < \frac{1+\chi^2}{2 \chi}$. 
\end{itemize}

In Fig.~\ref{fig-phase-dia} the different phases are plotted in the phase diagram of the extended Agassi. The first one, represented by a red sphere, corresponds to the spherical solution $(\varphi=0, \beta=0)$ and it is limited to the area with $\chi<1$,  $\Sigma<1$, and  $\Lambda<1$. The value of the energy in this region is $E/(j\varepsilon)=-1$. The limits of the region correspond to QPT surfaces, i.e., second order QPT for the vertical planes with  $\chi=1$ and $\Sigma=1$ (in red) and first order QPT for the plane $\Lambda=1$ (in green). The second area, represented by a blue oval, corresponds to the HF deformed solution $(\cos\varphi=\frac{1}{\chi},\beta=0)$ and it is limited by the surfaces $\chi=1$ (in red), which implies a second order QPT, $\chi=\Sigma$ (in blue), which supposes a first order QPT, and $\Lambda=\frac{1+\chi^2}{2\chi}$ (in green) that also implies a first order QPT. The value of the energy in this case is $E/(j\varepsilon)=-\frac{1+\chi^2}{2\chi}$. The third area, represented by a black oval, corresponds to the BCS deformed solution ($\varphi=0$, $\cos\beta=\frac{1}{\Sigma}$) and the value of the energy is $E/(j\varepsilon)=-\frac{1+\Sigma^2}{2\Sigma}$. The region is limited by the plane $\Sigma=1$ (in red),  corresponding to a second order QPT, and the green surface $\Lambda=\frac{1+\Sigma^2}{2\Sigma}$ and the plane $\chi=\Sigma$ (in blue) that correspond to first order QPTs. Finally, the fourth area, represented by the green crossed ovals, corresponds to the HF-BCS deformed solution $(\varphi=\frac{\pi}{2},\beta=\frac{\pi}{2})$ with energy $E/(j\varepsilon)=-\Lambda$. This region is limited by the green surface which implies a first order QPT. Also note that for the surface $\chi=\Sigma$ the closed valley solution ($\cos\varphi \cos \beta= 1/\chi$), represented by a thick ellipse, is also valid.

Take into account that the phases represented in the phase diagram are the ones corresponding to the deepest absolute minimum of the mean-field energy (including both surfaces A and B). However, in each region several phases can coexist, up to three (see \cite{Garc18} for details). In addition, in the line $\chi = \Sigma$ with $\Lambda=\frac{1+\chi^2}{2\chi}$,  four phases, HF, BCS, HF-BCS and the closed valley solutions, are degenerated. Finally, in the single point, $\chi=\Sigma=\Lambda=1$, the five solutions for the system  are degenerated. This provides with a richer phase diagram than for other studied complex systems, such as the two-fluid LGM model \cite{Garc16}, the proton-neutron IBM \cite{Aria04}, or for Hamiltonians with up two three-body interactions \cite{Fort11}.

\section{Comparison between exact and mean-field results: the large size limit} 
\label{sec-large}

A very convenient test of the kind of mean-field calculations presented so far is its comparison with the exact results. However, any exact diagonalization will be done for a finite size system, therefore finite size discrepancies between the exact and the mean-field results will always appear. These, of course, will be reduced as the size of the system increases, although they will always reach a limit due to the computational limitations.   
\begin{figure}[hbt]
  \centering
\includegraphics[width=1\linewidth]{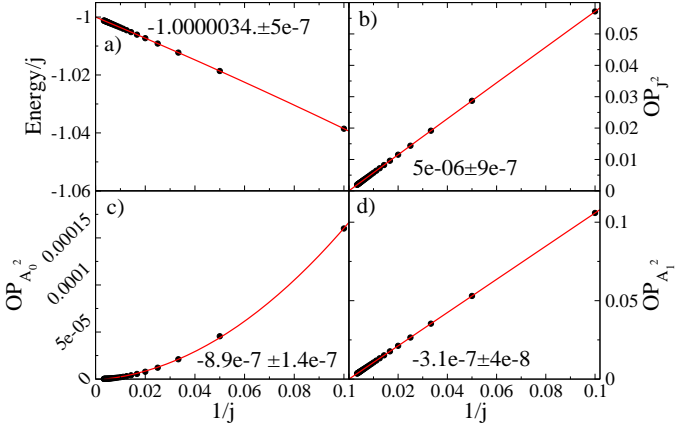}
\caption{1/j representation for exact calculations with $j$ ranging from $10$ to $300$. The parameters in the Hamiltonian are $\chi=0.5$,
  $\Sigma=0.5$, $\Lambda=0$. The numbers next to the red line stand for the thermodynamic value of the studied quantity (with its numerical error).  Panel a) corresponds to the ground state energy, panel b) to $OP_{J^2}$, panel c) to $OP_{A_0^2}$, and panel d) to $OP_{A_1^2}$ order parameters.} 
\label{fig-one_over-19}
\end{figure}

A possible comparison between the exact and the mean-field results is a $1/j$ analysis. This will provide the value of the considered quantity in the thermodynamic limit and, therefore, should coincide with the mean-field value. Here, the approach presented in \cite{Bert10} is closely followed, where the ground-state energy is written as,
\begin{equation}
  \frac{\mathcal{E}}{j}=a+\frac{b}{j}+\frac{c}{j^2}+\frac{d}{j^3}+O\left(\frac{1}{j^4}\right),
  \label{Ej}
\end{equation}
where it is assumed that the two-body coefficients are rescaled with the size of the system, i.e., they are multiplied by $1/j$. The value of the energy in the thermodynamic limit will correspond, therefore, to the intercept, $a$.
\begin{figure}[hbt]
  \centering
\includegraphics[width=1\linewidth]{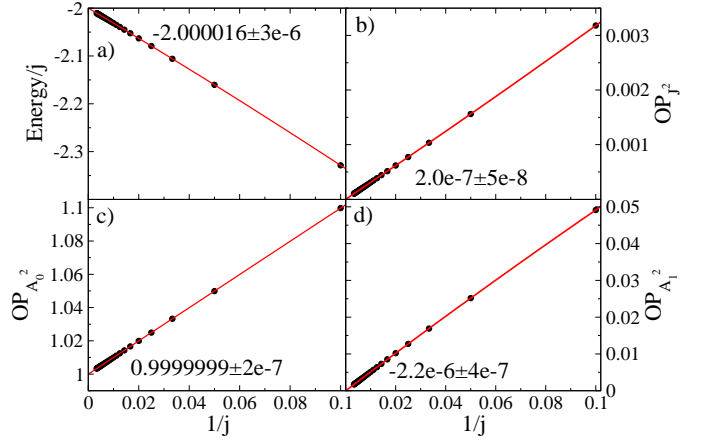}
\caption{Same caption as in Fig.~\ref{fig-one_over-19} but with parameters $\chi=0.5$, $\Sigma=0.5$, $\Lambda=2.$} 
\label{fig-one_over-22}
\end{figure}

Note that expression (\ref{Ej}) is, in principle, only valid well apart from the critical point where the observables behave as $j^\alpha$ in a second-order phase transition and the value of $\alpha$ depends on the system under study.  

This expansion is justified on the light of a shift transformation $a^\dag=b^\dag+\lambda\sqrt{j}$, because for $\frac{\mathcal{O}_1}{j}$, where $\mathcal{O}_1$ is a one-body term, the highest order in $j$ will be $j^0$, while the lowest $j^{-1}$. In the case of a two-body term, $\mathcal{O}_2$, appropriately scaled, $\frac{\mathcal{O}_2}{j^2}$, the highest order in  $j$ is, again,  $j^0$ while the lowest $j^{-2}$. In our calculations,  up to quadratic terms in the $1/j$ expansion of the studied observable will be considered.
\begin{figure}[hbt]
  \centering
\includegraphics[width=1\linewidth]{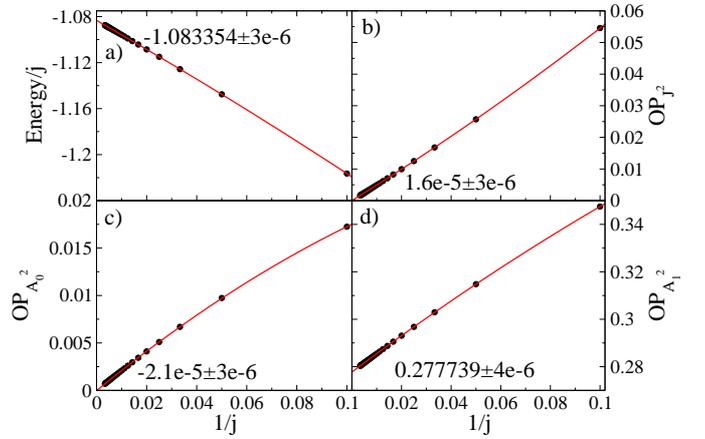}
\caption{Same caption as in Fig.\ref{fig-one_over-19} but with parameters $\chi=0.5$, $\Sigma=1.5$, $\Lambda=0.$}
\label{fig-one_over-25}
\end{figure}

In order to obtain the thermodynamic limit of the ground-state energy, its representation as a function of $1/j$ is plotted from $j=10$ to $j=300$ in steps of $j=10$. The value of the intercept, obtained after performing a quadratic least squares fit, will correspond to the thermodynamic limit of the quantity under study. In principle, this analysis can  be extended to any other excited state, although in this case it cannot be compared with the mean-field value.

The values of the effective order parameters are also compared using the same procedure. These are defined in terms of the expectation values of the following operators for the ground state,
\begin{eqnarray}
\label{ef_j} OP_{{J}^2}&=&\frac{\langle (J^+)^2 \rangle + \langle (J^-)^2\rangle}{2 j^2},\\
\label{ef_a0}OP_{{A}^2_0}&=&\frac{\langle A_0^+A_0\rangle}{j^2},\\
\label{ef_a2}OP_{{A}^2_1}&=&\frac{\langle A_1^+A_1\rangle +\langle A_{-1}^+A_{-1}\rangle}{2 j^2}.
\end{eqnarray}

We first study in Fig.~\ref{fig-one_over-19} the case $\chi=0.5$, $\Sigma=0.5$, $\Lambda=0$  (note that in all numerical calculations it is assumed $\varepsilon=1$), which corresponds to a point in the phase diagram with $\varphi=0$, $\beta=0$, and $E_{gs}/j=-1$. The value  of the energy in the thermodynamic limit is $E_{gs}/j=-1.0000034 \pm 5\times 10^{-7}$ to be compared with $-1$. In the case of the order parameters, their obtained values in the thermodynamic limit are $(-8.9 \pm 1.4) \times 10^{-7}$, $(-3.1 \pm 0.4) \times 10^{-7}$ and $(5.0 \pm 0.9) \times 10^{-6}$ for $OP_{{A}^2_0}$, $OP_{{A}^2_1}$, and $OP_{{J}^2}$, respectively. These have to be compared with the mean-field values that are zero for the three cases. This supposes a difference between the mean-field and the extracted thermodynamic limit values of the order of $10^{-6}$ or smaller for all the magnitudes under study.

In Fig.~\ref{fig-one_over-22} the case $\chi=0.5$, $\Sigma=0.5$, $\Lambda=2$ is considered. It corresponds to a point in the phase diagram with $\varphi=\pi/2$, $\beta=\pi/2$, and $E_{gs}=-\Lambda$. The  value of the energy in the thermodynamic limit is $E_{gs}/j=-2.000016 \pm 3\times 10^{-6}$ to be compared with the mean field value $-2$. In the case of the order parameters, their values in the thermodynamic limit result to be $0.9999999 \pm 2\times 10^{-7}$,  $(-2.2 \pm 0.4) \times 10^{-6}$, and  $(2.0 \pm 0.5) \times 10^{-7}$ for $OP_{{A}^2_0}$, $OP_{{A}^2_1}$, and $OP_{{J}^2}$, respectively. These have to be compared with the mean-field values that are $1$, $0$, and $0$, respectively. This supposes a difference between the mean-field and the values in the thermodynamic limit smaller than $10^{-7}$ for the energy and of the order of $10^{-6}$ for the order parameters.
\begin{figure}[hbt] \centering
\begin{tabular}{c}
\includegraphics[width=1\linewidth]{energ-order-par-15ll-quad-error.eps} \\
\includegraphics[width=.3\linewidth]{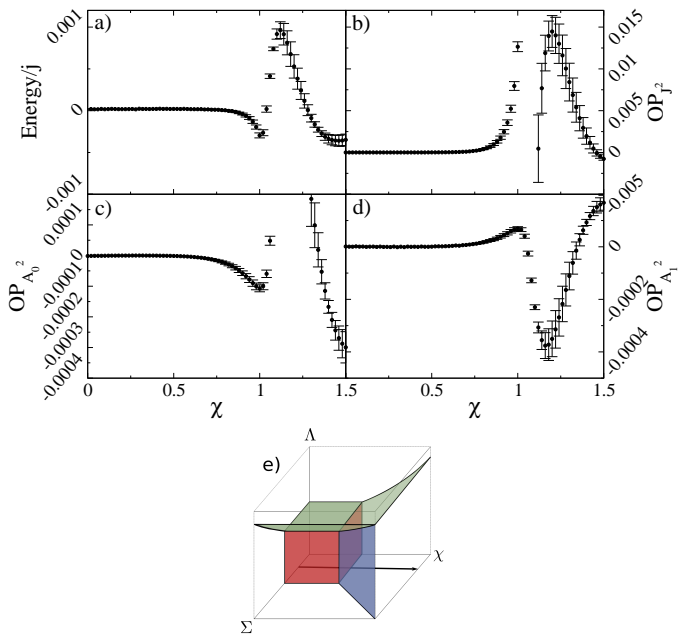}
\end{tabular}
\caption{Difference between the thermodynamic value and the mean-field one, including its error, for the ground-state energy per fermion pair and the order parameter values, with Hamiltonian parameters $\Sigma=0.5$, $\Lambda=0$, as a function of $\chi$. Panel a) corresponds to the ground state energy, panel b) to $OP_{J^2}$, panel c) to $OP_{A_0^2}$, panel d) to $OP_{A_1^2}$ order parameters, and panel e) to the schematic representation of the trajectory in the parameter space.} 
\label{energ-order-par-15}
\end{figure}

In Fig.~\ref{fig-one_over-25} the case $\chi=0.5$, $\Sigma=1.5$, $\Lambda=0$ is selected. This corresponds to a point in the phase diagram with $\varphi=0$, $\beta=\arccos(1/\Sigma)$, and $E_{gs}=-\frac{\Sigma^2+1}{2\Sigma}$. The  value of the energy obtained for the thermodynamic limit is $E_{gs}/j=-1.083354 \pm 3\times 10^{-6}$ to be compared with the mean-field value $-1.083333$ $(-13/12)$. In the case of the order parameters their values in the thermodynamic limit read as $(-2.1 \pm 0.3) \times 10^{-5}$,  $0.277739 \pm 4\times 10^{-6}$, and  $(1.6 \pm 0.3)\times 10^{-5}$ for $OP_{{A}^2_0}$, $OP_{{A}^2_1}$, and $OP_{{J}^2}$, respectively. These results are to be compared with the mean-field values that are $0$, $5/18\approx 0.277778$, and $0$ respectively. This supposes, once more, a difference between the mean-field and the extracted thermodynamic limit of the order of $10^{-5}$ for the energy and for the order parameters.


Once the technique has been presented, the comparison between the
mean-field and the exact results in the large-$j$ limit for selected
trajectories is presented now. For specific values of the control
parameters, a calculation with $j$ ranging from $10$ to $200$ is
performed extracting from a least squares fit to a function
$a+\frac{b}{j}+\frac{c}{j^2}$ the value of the intercept, $a$, and its
corresponding error. This value corresponds to the thermodynamic limit
of the quantity under study. Same quantities as in the preceding
discussion are studied, but to better appreciate the agreement between
both, the mean-field and the exact results in the thermodynamic limit,
the difference between both values is plotted in the following
figures, together with the value of the statistical error extracted
from the least squares fit. The selected trajectories move between two
different phases crossing a QPT surface either of first or of second order. These trajectories are those studied in \cite{Garc18}, although in the later case the exact calculations were performed for $j=50$.
\begin{figure}[hbt]
  \centering
\begin{tabular}{c}
\includegraphics[width=1\linewidth]{energ-order-par-10ll-quad-error.eps} \\
\includegraphics[width=.3\linewidth]{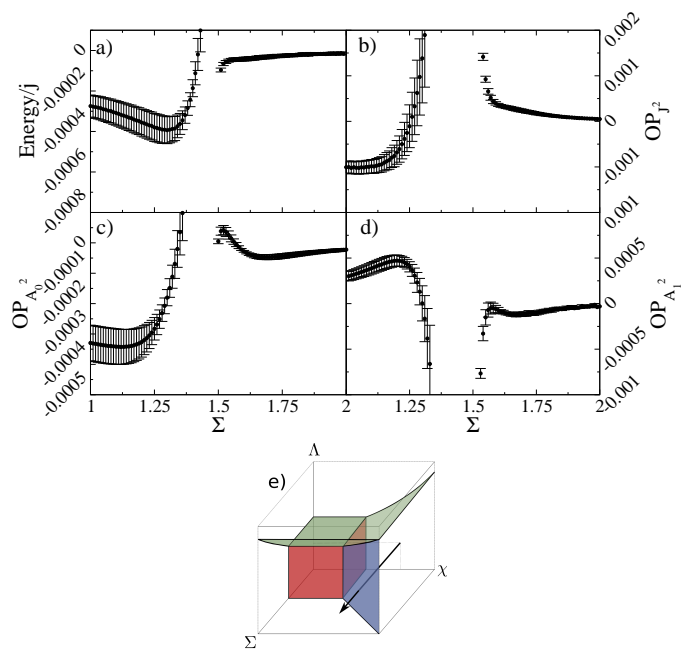}
\end{tabular}
\caption{Same caption as in Fig.~\ref{energ-order-par-15} but for $\chi=1.5$, $\Lambda=0.5$, as a function of $\Sigma$.}
\label{energ-order-par-10}
\end{figure}

As a first case, the trajectory  $\Sigma=0.5$,  $\Lambda=0$ as a function of $\chi$ is presented in Fig.~\ref{energ-order-par-15}. As shown in panel e), this trajectory crosses a second order QPT at $\chi=1$. The values for the difference between the large-$j$ thermodynamic limit and the mean-field values for the ground state energy (panel a), and the order parameters associated to the expection values of $J^2$ (panel b), $A_0^2$(panel c), and $A_1^2$ (panel d), including their errors bars, are plotted. A remarkable agreement between the values in the thermodynamic limit and the mean-field ones is observed, although the agreement is clearly better in the spherical phase than in the HF deformed one. Moreover, the presented approach, as already mentioned, is not well suited for the region around the QPT, that is around $\chi\approx 1$, where the agreement worsen. Note that some points around the QPT may lay out of the plotted scale.

The second trajectory studied is $\chi=1.5$, $\Lambda=0.5$, as a function of $\Sigma$ and the results are displayed in Fig.~\ref{energ-order-par-10}. As shown in panel e), this trajectory crosses a first order QPT at $\Sigma=1.5$. As in the preceding case, a remarkable agreement between the values in the thermodynamic limit and the mean-field ones is observed, although here the agreement is better for the BCS-deformed phase than for the HF-deformed region. Again, larger discrepancies, as expected, are observed in a small region close to the critical point at $\Sigma=1.5$. Note that some points around the QPT may lay out of the plotted scale.

\begin{figure}[hbt]
  \centering
\begin{tabular}{c}
\includegraphics[width=1\linewidth]{energ-order-par-12ll-quad-error.eps} \\
\includegraphics[width=.3\linewidth]{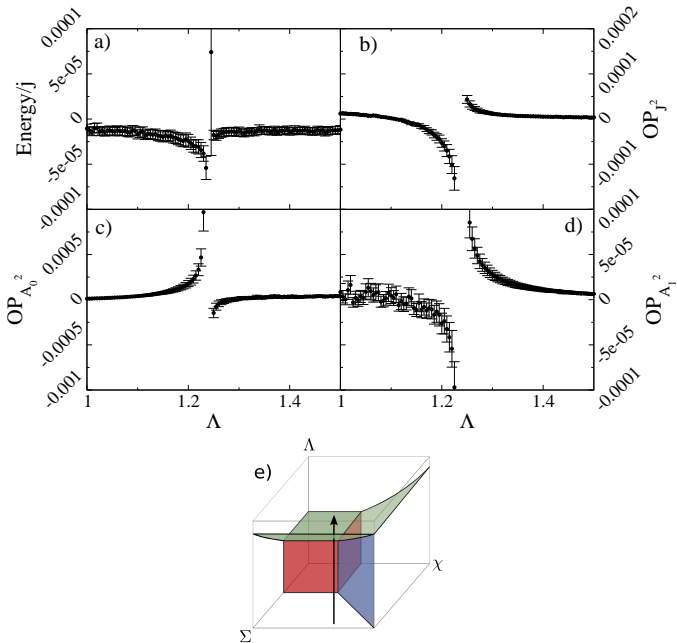}
\end{tabular}
\caption{Same caption as in Fig.~\ref{energ-order-par-15} but for $\chi=1.5$, $\Sigma=2$, as a function of $\Lambda$.}
\label{energ-order-par-12}
\end{figure}

The last trajectory analysed is $\chi=1.5$, $\Sigma=2$, as a function of $\Lambda$ and the results are plotted in Fig.~\ref{energ-order-par-12}. As shown in panel e), this trajectory crosses a first order QPT at $\Lambda=5/4$. In this case the agreement is also noticeable. The difference in all studied magnitudes between the mean-field results and those obtained in the large-$j$ thermodynamic limit is basically zero except in a small region around the critical point $\Lambda=5/4$. Note that some points around the QPT may lay out of the plotted scale.

\section{Summary and conclusions}
\label{sec-conclusions}
An extended version of the Agassi model that includes an extra $A_0^\dag A_0$ contribution (a non-standard pairing contribution) in the Hamiltonian has been reviewed. Therefore, the model depends on four free control parameters, $\varepsilon$, $g$, $V$, and $h$. However, a non-vanishing value for $\varepsilon$ has always been considered. Hence, the number of effective free parameters is three: $V=\frac{\varepsilon\chi}{2j-1}$, $g=\frac{\varepsilon\Sigma}{2j-1}$, and $h=\frac{\varepsilon\Lambda}{2j-1}$. The underlying algebra of the model, o(5), has been presented, and both the fermionic and the Schwinger boson representation of the model have been worked out. A mean-field approach for the extended Agassi model through a HFB mean-field approach is presented. In addition, the alternative use of a condensate of bosons in the Schwinger representation is discussed. The corresponding energy surfaces are obtained and the phase diagram of the model is established. The phase diagram presents four different regions: spherical, HF deformed, BCS deformed, and combined HF-BCS deformed phases. Moreover, there is a surface in which a special solution, called closed valley, exists.  These phases are separated by several surfaces corresponding either to first or second order QPTs.  In addition, there is a line in which four phases coexist (HF deformed, BCS deformed, combined HF-BCS deformed, and close valley deformed minimum) and are degenerated. There exists a single point $\chi=\Sigma=\Lambda=1$ in which the five phases are degenerated (spherical, HF deformed, BCS deformed, combined HF-BCS deformed, and close valley deformed minimum).

Finally, the exact results in the large-$j$ limit of the model are extracted by using a quadratic $1/j$ expanssion and compared with the mean-field results. First, the exact large-$j$ limit is obtained for specific values of the control parameters performing a $1/j$ analysis with $j$ ranging from $j=10$ to $j=300$. Second, we have applied the technique for selected trajectories that cross QPT regions. In all cases, the large-$j$ analysis provides very accurate values for the energy and the order parameters, and the correspondence with the mean-field values is noticeable. Therefore, the proposed technique seems to be suitable to calculate the thermodynamic limit of any other observables. 


\begin{ack}
This work has been supported by the Spanish Ministerio de Econom\'{\i}a y Competitividad and the European regional development fund (FEDER) under Projects No. FIS2017-88410-88410-P, FIS2014-53448-C2-2-P and FIS2015-63770-P, and by Consejer\'{\i}a de Econom\'{\i}a, Innovaci\'on, Ciencia y Empleo de la Junta de Andaluc\'{\i}a (Spain) under Group FQM-160 and FQM-370.
\end{ack}


\end{document}